# A plausible interpretation of the density scaling
# of the diffusivity in viscous liquids

**Anthony N. Papathanassiou ***

*University of Athens, Physics Department, Solid State Physics Section,*

*Panepistimiopolis, 157 84 Zografos, Greece*

Fundamental thermodynamic concepts and an earlier elastic solid-state point defect model are employed to formulate an analytical second-order polynomial function describing the density scaling of the diffusion coefficient in viscous liquids. The scaling exponent is correlated, within the approximations made in the present approach, with the pressure derivative of the isothermal bulk modulus. Our findings are compared with computer simulation results.

**Keywords :** Viscous liquids ; diffusivity ; pressure; scaling

(*)   E-mail address: antpapa@phys.uoa.gr

## 1. Introduction

Viscous liquids exhibit extraordinary values of viscosity compared with those of ordinary liquids. In the extreme viscosity limit (i.e., close to the calorimetric glass-transition) molecular transport is retarded and most molecular motion is vibrational [1] and the viscous liquid resembles a disordered solid [2] that flows [1, 3]. A series of review articles on the properties of glass forming liquids were published recently [1, 2, 3, 4]. Ultra-viscous matter has exceptional features [5, 6] and universalities which are not well understood yet [2], such as the strongly non-Arrhenius temperature dependence [7] of the structural relaxation time and the strong temperature dependence of the activation energy of the so-called fragile glass formers [1]. A dynamic quantity $\chi$, such as structural relaxation time $\tau$, viscosity $\eta$ or diffusion coefficient D in viscous liquids seems to scale with some fundamental quantities like density $\rho$ and temperature. A common scaling expression found in the literature [4, 8, 9, 10, 11] is:

$$\chi = F(\rho^{\gamma} / T) \tag{1}$$

where $\rho$ denotes the density, $\gamma$ is a scaling exponent, T is the temperature and F is a (scaling) function, which is a priori unknown. Most of the experimental evidence for thermodynamic scaling is for the structural relaxation time and viscosity. Deviations from inverse proportionality between D and $\tau$ occur on approaching the glass transition, whereas enhanced translation relative to reorientation occurs, so scaling of $\tau$ doesn't guarantee scaling of D. However, these deviations from Stokes-Einstein may be small enough to not be apparent in a plot of super positioned data. Or maybe they are subsumed in a small change in $\gamma$ [12]. The correlation of the scaling exponent $\gamma$, which is a material constant, with the physical properties of the viscous state is a matter of ongoing exploration. Computer simulations of Lennard-Jones liquids, with the exponent of the repulsive term taking the values 8, 12, 24 and 37, revealed that density scaling is valid and the exponent $\gamma$ is roughly one third of the exponent of the effective inverse power repulsive term [13]. Molecular dynamics also indicated that strong virial/potential-energy correlations also reflect the effective inverse power law and scaling occurs in strongly correlating viscous liquids [14]. Recent progress on the role of thermodynamic elastic models to the density scaling of the diffusivity appeared recently [15]. On the other hand, following the Avramov entropy model [16] for the structural relaxation time, $\gamma$ was correlated with the thermodynamic Grüneisen parameter $\gamma_G$ [4, 11, 17, 18].

Solid-state elastic models seem to play a prominent role in describing these phenomena. The distinctive role of thermodynamic point defect models for understanding the viscous state was mentioned recently [19, 15]. In the present work, we start from thermodynamic concepts and, by using elastic point defect models and provide an analytical equation governing the density scaling of the diffusion coefficient in viscous liquids [15]. The morphology of the scaling function agrees with up to date experimental results and computer simulations. The present formulation predicts that the scaling function is practically pressure insensitive, in agreement with recent computer simulations of binary Lennard-Jones systems, for various exponent values of the repulsive term of the potential results [13].

## 2. Theoretical formulation

Isotherms of the logarithm of the relaxation time of viscous liquids as a function of pressure have a clear non-linear behavior [20, 21, 22, 23]. The pressure dependence of logarithm of the diffusion coefficient obtained from molecular dynamics simulations [24] deviates from linearity, as well. $\ln D$ vs pressure shows a downward curvature. The increase of the (absolute) value of slope of the latter curve with pressure was speculatively interpreted, as a change in the transport mechanism in viscous liquids, occurring at pressure where hopping of particles become noticeable [24]. Alternatively, it was attributed [24], according to the free-volume theory, to random close packing occurring at elevated pressure. However, the curvature in diffusivity - pressure plots was interpreted earlier: Varotsos and Alexopoulos suggested that such curvature results from a pressure dependent activation volume [25]; if $g^{act}$ denotes the Gibbs free energy for diffusion, the corresponding activation volume is defined as $\upsilon^{act} \equiv \left(\partial g^{act}/\partial P\right)_T$. Since there is no physical argument to regard $\upsilon^{act}$ as constant, the compressibility of the activation volume is defined as $\kappa_T^{act} \equiv -\left(\partial \ln \upsilon^{act}/\partial P\right)_T$ [25]. The latter can be positive, negative or zero. The data reported in Ref. [24] indicate that $\kappa_T^{act} < 0$ for viscous liquids and, to a first approximation, we regard $\kappa_T^{act}$ constant. The isothermal pressure evolution of the reduced diffusion coefficient $D^*(P)$ (i.e., the diffusion coefficient normalized by the zero-pressure diffusion coefficient) can be approximated by the following analytical equation [25]:

$$\ln D^*(P) = -\left[\frac{\upsilon_0^{act}}{kT} - \frac{\gamma_G}{B_0}\right]P + \left[\frac{\upsilon_0^{act}\kappa_T^{act}}{2kT}\right]P^2 \qquad (2)$$

where $\upsilon_0^{act}$ and $B_0$ denote the zero (ambient) pressure activation volume and isothermal bulk modulus, respectively. It is evident that, whenever $\kappa_T^{act}$ is zero (i.e., $\upsilon^{act}$ is constant), Eq. (2) reduces to a simple well-known linear relation. From another viewpoint, the curvature may be interpreted if $\upsilon^{act}$ is not single-valued, but obeys a normal distribution [26, 27].

Solid-state thermodynamic elastic point defect models suggest that the activation volume is proportional to the activation Gibbs free energy $g^{act}$: According to the so-called $cB\Omega$ model [28, 29, 30, 31]:

$$g^{act} = cB\Omega \qquad (3)$$

where c is a roughly constant and $\Omega$ is a volume related with the mean atomic volume. Differentiating Eq. (3) with respect to pressure we get:

$$\upsilon^{act} = B^{-1}\left[(\partial B/\partial P)_T - 1\right]g^{act} \qquad (4)$$

In the viscous state, the activation enthalpy is usually a few tenths of kT (or, more) [2, 10]. We can write $h^{act} \approx \Lambda kT$, where $\Lambda$ is a material's constant that is a function of temperature, in general and usually takes values of the order of 10 [10]' The activation entropy $s^{act}$ is usually only about a few k, thus, $g^{act} = h^{act} - Ts^{act}$ is of the same order of magnitude as $h^{act}$. Subsequently, at zero pressure, Eq. (4) is rewritten as:

$$\upsilon_0^{act} \approx \frac{\Lambda}{B_0}[(\partial B/\partial P)_T - 1]kT \qquad (5)$$

Assuming that $\left|\kappa_T^{act}\right| \approx \frac{1}{B_0}$ [32], Eq. (2), when combined with Eq. (5), yields:

$$\ln D^*(P) = -\left[\frac{\Lambda}{B_0}[(\partial B/\partial P)_T - 1] - \frac{\gamma_G}{B_0}\right]P - \frac{\Lambda}{2B_0}[(\partial B/\partial P_T) - 1]P^2 \qquad (6)$$

The latter equation can explicitly be expressed as a function of (reduced) density: By definition, the isothermal bulk modulus is $B \equiv -(\partial P/\partial \ln V)_T$. Recalling that $\rho \equiv m/V$, we get $B = (\partial P/\partial \ln \rho)_T$. To a first approximation, we employ the well-known Murnaghan equation of state, which implies that the isothermal bulk modulus increases linearly with pressure: i.e., $B(P) = B_0 + (\partial B/\partial P)_T P$, where $(\partial B/\partial P)_T$ is assumed to be roughly constant. Volumetric data of various viscous liquids confirm that the latter $B(P)$ function is a fairly good approximation [33]. Under the constrain of a linear $B(P)$ relation, the solution of the differential equation $B = (\partial P/\partial \ln \rho)_T$, is [34]:

$$\rho^{(\partial B/\partial P)_T} = 1 + \frac{(\partial B/\partial P)_T}{B_0}P \qquad (7)$$

Eq. (7) permits the alteration of the variable $P$ to $\rho$ appearing in the reduced diffusivity (Eq. (6)):

$$\begin{aligned}
\ln D^*(\rho^\gamma) = &-\frac{\Lambda\gamma_G}{(\partial B/\partial P)_T^2}(\rho^\gamma)^2 \\
&-\frac{\gamma_G}{(\partial B/\partial P)_T}\left[2\Lambda\left(1 - \frac{1}{(\partial B/\partial P)_T}\right) - 1\right]\rho^\gamma \\
&+\frac{\gamma_G}{(\partial B/\partial P)_T}\left[\Lambda\left(2 - \frac{1}{(\partial B/\partial P)_T}\right) - 1\right]
\end{aligned} \qquad (8)$$

where $\gamma \equiv (\partial B / \partial P)_T$. We stress that the later identification is constrained by the approximations, assumptions and restrictions asserted in the present work. Further progress provided a more refined approach [35].

## 3. Results and Discussion

We mention that Eq. (8) captures the interconnection of diffusion parameters with elastic properties of the material (within the frame of the $cB\Omega$ elastic solid state point defect model) and the universal feature of glass-formers that the activation enthalpy is usually roughly tenths of kT (i.e., $h^{act} \approx \Lambda kT$, where $\Lambda$ is a function of temperature taking values of the order of ten). Moreover, Eq. (8) provides a direct connection between the scaling exponent $\gamma$ and $(\partial B / \partial P)_T$, *under the assumptions and approximations made in the present work* (a couple of potential assumptions are that the diffusivity Eq. (2) is applicable in the ultra-viscous state and $g^{act}$ is proportional to the bulk modulus B). Further work can improve the validity of Eq. (8) by including the temperature dependence of the activation enthalpy, which does it differently in different materials [10]. The diffusivity scaling equation predicts that:

*(i)* The (natural) logarithm of the reduced diffusion coefficient $D^*$ is a decreasing function of $\rho^\gamma$.

*(ii)* The function $\ln D^*(\rho^\gamma)$ is a second order polynomial with downward curvature. The latter form, which is based on physical arguments, is suitable to fit isothermal density scaling diffusion data, instead of using arbitrary equations [36].

*(iii)* The slope of the $\ln D^*(\rho^\gamma)$ curve depends on $\Lambda$, $\gamma_G$, and $(\partial B / \partial P)_T$ which are characteristic physical quantities of the viscous liquid.

*(iv)* Different $\ln D^*(\rho^\gamma)$ isotherms obtained at different pressures for the same viscous liquid, collapse on a unique master curve. This is due to the fact that $\Lambda$ and $\gamma$ are constant for the viscous liquid under study. The present formalism gives the theoretical interpretation of computer simulation results of Lennard-Jones liquids m-6 ($8 \leq m \leq 36$) in normal and moderately super-cooled states [13], which indicated that the diffusion coefficient plotted against $\rho^\gamma/T$ at different pressures, accumulate on a single curve [37].

The density and temperature scaling of dynamic properties of viscous liquids is relatively a recent speculation [8]. At present, apart from numerical simulations, experimental work on density and temperature scaling is available for the structural relaxation time and the viscosity, but missing for the diffusivity. Only numerical results are available from important groups, which make predictions on the scaling of diffusivity; the diffusivity scaling exponents predicted are spanning over a broad range, from 3.5 to 13.7 [13, 36, 38]. Phenylphyhalein-dimethylether (PDE), which is a typical viscous liquid, has a pressure derivative of the bulk modulus $(\partial B / \partial P)_T = 9.76$ at 372.6 K [33], which yields (according to the present work) $\gamma = 9.76$. This value lies within the range of the above-mentioned predictions from computer simulations. Concerning the diffusivity, it seems that we are at a stage that simulations and theory are temporarily advancing in relation with the experimental work. The currently published simulations and the present theoretical work (initiated by our earlier publication [15]) exhibit the emerging necessity of investigating experimentally the density and temperature scaling of diffusion coefficient in viscous liquids.

## 4. Conclusion

The derivation of Eq. (8), which was based on thermodynamic concepts and the cBΩ elastic solid-state point defect model, confirms the statement of Dyre [6] that viscous flow events can be correlated with defect motion in crystals: free energies from activation for self-diffusion are proportional to the isothermal bulk modulus (cBΩ model) and, if shear and bulk moduli are proportional to their temperature and pressure variation, then the cBΩ model becomes equivalent to the shoving model [6], which is based on the fact that activation energy is dominated by the work done to shove aside the surroundings [2, 39].

## Acknowledgements

The author is grateful to D. Coslovich (Wien Technical University), C.M. Roland (Naval Research Laboratory, USA) and I. Sakellis (Athens University) for helpful recommendations. An invitation from Mike Roland to give an oral presentation of this work is greatly acknowledged. Participation to the Conference was granted by a Specific Research Fund, University of Athens (Project Kapodistrias).

when plotted on the same diagram, they superposition each other and can hardly be distinguished).